\documentclass[preprint2]{aastex}

\usepackage{graphicx}
\usepackage{ulem}
\usepackage{amsmath}
\usepackage{wrapfig}

\shorttitle{Long-term Evolution of Direct Impact DWDs}
\shortauthors{Kremer, Sepinsky, and Kalogera}

\begin{document}

\title{Long-term evolution of double white dwarf binaries accreting 
through direct impact}

\author{Kyle Kremer}
\affil{Center for Interdisciplinary Exploration and Research in Astrophysics (CIERA)\\
Department of Physics and Astronomy\\
Northwestern University, 2145 Sheridan Road, Evanston, IL 60208}
\email{kremer@u.northwestern.edu}

\and

\author{Jeremy Sepinsky}
\affil{Department of Physics and Electrical Engineering, The University 
of Scranton\\ Scranton, PA 18510}
\email{jeremy.sepinsky@scranton.edu}

\and

\author{Vassiliki Kalogera}
\affil{Center for Interdisciplinary Exploration and Research in Astrophysics (CIERA)\\
Department of Physics and Astronomy\\
Northwestern University, 2145 Sheridan Road, Evanston, IL 60208}
\email{vicky@northwestern.edu}

\begin{abstract}

We calculate the long-term evolution of angular momentum in double white
dwarf binaries undergoing direct impact accretion over a broad range of
parameter space. We allow the rotation rate of both components to vary,
and account for the exchange of angular momentum between the spins of the
white dwarfs and the orbit, while conserving the total angular momentum.
We include gravitational, tidal, and mass transfer effects in the
orbital evolution, and allow the Roche radius of the donor star to vary
with both the stellar mass and the rotation rate. We examine the
long-term stability of these systems, focusing in particular on those
systems that may be progenitors of AM CVn or Type Ia Supernovae. We find
that our analysis yields an increase in the predicted number of stable
systems compared to that in previous studies. Additionally, we find that
by properly accounting for the effects of asynchronism between the donor
and the orbit on the Roche-lobe size, we eliminate oscillations in the
orbital parameters which are found in previous studies. Removing these
oscillations can reduce the peak mass transfer rate in some systems,
keeping them from entering an unstable mass transfer phase.

 \end{abstract} 


\maketitle

\clearpage

\section{Introduction}

Binary systems containing compact objects are of great interest in a 
variety of areas in astrophysics. Of particular importance are double 
white dwarf (DWD) binary systems where both components are white dwarfs, 
which may make up the largest fraction of close binary stars 
\citep{Marsh1995}.

Following common envelope evolution, DWD binaries may emerge with a 
sufficiently small semi-major axis, allowing gravitational radiation to 
drive the stars closer together on an astrophysically interesting time 
scale. These close DWD systems are of prime importance as low-frequency 
gravitational-wave sources \citep{Hils1990,Hils2000,Nelemans2001} as 
well as progenitors of Type Ia Supernovae \citep{Maoz2014}.

As energy loss due to gravitational waves drives the degenerate 
components of a DWD binary together, it is possible for the system to 
enter into a stable semi-detached state, in which the less massive 
component will fill its Roche lobe and begin transferring matter to its 
companion. Such systems may result in the formation of AM CVn 
\citep{Nather1981,Tutukov1996,Nelemans2001}, which 
have extremely short orbital periods. During mass 
transfer, a stream of matter is pulled from the donor star through the 
inner Lagrangian point.  If the matter stream does not impact the 
surface of the companion star, then the mass lost from the donor is 
expected to settle into a disc \citep{Frank2002}.  Torques exerted 
between the discs and the component white dwarfs allow angular momentum 
stored in the disc to be transferred back to the orbit 
\citep{Soberman1997,Frank2002}. This allows matter in the disc to fall 
onto the companion star while slowing the orbit contraction, 
increasing the potential for a stable, long-lived binary system.  \footnote{We note that we have not considered the possibility of the white-dwarf components having some residual, non-degenerate, hydrogen-rich outer layers. In a recent analysis, \citet{Shen2015} showed that the inclusion of such effects and associated nova-like outbursts may predominantly lead to mergers for both direct-impact and disc accretion.}

For some DWD binaries, the mass transfer stream will directly impact the 
surface of the companion star.  In doing so, there is no longer an 
obvious mechanism to return the orbital angular momentum from the 
transferred mass to the orbit.  The division between the disc accretion 
and direct impact has been studied before \citep[see][and references 
therein, hereafter Paper I, Marsh2004, and Gokhale2007, 
respectively]{Sepinsky2013, Marsh2004, Gokhale2007}.

As in Marsh2004 and Gokhale2007, we showed in Paper I that mass transfer through direct impact can either increase or 
decrease the semi-major axis of the system, depending on the amount of 
angular momentum and mass exchanged between the components.  If direct 
impact drives the system apart, both Marsh2004 and Gokhale2007 showed 
that a stabilizing accretion disc is likely to be created.  If 
direct-impact mass transfer decreases the semi-major axis, the mass 
transfer rate may eventually become unstable, which can result in a 
merger.  If the total mass of the system is in excess of the 
Chandrasekhar limit, such merger events could lead to a Type Ia 
supernova \citep{Woosley1986}.

Marsh2004 and Gokhale2007 both calculated of the long term evolution of 
such systems. Marsh2004 concluded that the population of DWD binaries is 
likely lower than previously anticipated as a result of the high 
percentage of systems undergoing direct-impact accretion that become 
unstable. Gokhale2007 improved upon the analysis of Marsh2004 by 
permitting the spin of the donor to vary and by including the 
effects of tidal forces from the donor star (Marsh2004 only examined 
tidal forces arising from asynchronicity between the accretor and the 
orbit. In their calculation, the spin of the donor was fixed to that of the orbit throughout so that no tides arose from asynchronicity between the donor and orbit.) The modifications of 
Gokhale2007 resulted in an increase in the number of stable systems which can be seen in Figures 
2 and 3 in section 4. In Paper I we built upon both of these 
studies, providing direct ballistic integrations of the mass transfer 
stream (as opposed to the previous method using an 
approximation adopted from \citet{Verbunt1988}).  We showed that 
removing this approximation which 
accounts for the complex three-body dynamics of the 
ejected mass can have a significant impact upon the 
angular momentum exchange.

In this paper, we apply the results of Paper I to determine the 
long-term evolution of double white dwarf binaries in a fully 
self-consisent manner.  In section 2, we introduce the equations which 
govern the long-term evolution of DWD systems and discuss the 
differences between this method and the method utilized by 
previous studies. In section 3 we discuss some details of our 
long-term numerical integrations. In section 4 we discuss the 
results of our solutions and analyze the results in comparison to those 
of previous works.  We conclude in section 5.

\section{Equations of Long-term Evolution}

\subsection{Basic Assumptions}

Following Paper I, we consider a close binary system of two white dwarfs 
with masses $M_A$ and $M_D$, volume-equivalent radii of $R_A$ and $R_D$, 
and uniform rotation rates $\Omega_A$ and $\Omega_D$ with axes 
perpendicular to the orbital plane for the accretor and donor, 
respectively. \footnote{Throughout this paper, the subscripts ``A" and 
``D" will correspond to the accretor and donor, respectively.} We assume 
the mass of each star is distributed spherically symmetrically. As in 
the analyses of Marsh2004 and Gokhale2007, we assume the binary to 
remain in a circular Keplarian orbit throughout its evolution. The 
radius of each object is assigned following Eggleton's zero-temperature 
mass-radius relation (equation 15 of \citet{Verbunt1988}). We assume 
that both the donor and accretor initially rotate sychronously with the 
orbit (Marsh2004, Gokhale2007).\footnote{We note that there may 
exist a substantial population of DWD with non-zero eccentricity and 
asynchronous components prior to the onset of direct impact accretion 
(\citet{Willems2007, Bours2014} and references 
therein). For this 
analysis, we limit our scope to systems that are initialy circular.  
Initially circular systems are expected to remain circular due to tidal 
circularization.  A more thorough treatment of the initial 
parameter-space including eccentricity and 
asynchronicity will be presented in a forthcoming 
paper.} We choose the initial semi-major axis of the orbit such that the 
volume-equivalent radius of the donor ($R_D$) is equal to the 
volume-equivalent radius of its Roche lobe as fit by 
\citet{Eggleton1983}.

\subsection{Evolution of angular momentum}
\label{sec:angmom}

As described in Marsh2004 and Gokhale2007, the orbital evolution of a 
circular binary system can be described by its orbital angular momentum.  
We begin by presenting each of the components that can lead to a change 
in the orbital angular momentum.

The total angular momentum, $J_{tot}$, of a binary system is given by 
the sum of the orbital angular momentum, $J_{orb}$, and the spin angular 
momenta, $J_{spin,A}$ and $J_{spin,D}$, of the
accretor and donor, respectively:
\begin{equation}
\label{eq-1}
J_{tot} = J_{orb} + J_{spin,A} + J_{spin,D}.
\end{equation}

The orbital angular momentum is given by
\begin{equation}
\label{eq-2}
J_{orb} = \sqrt{\frac{Ga}{M}}M_AM_D
\end{equation}
where $M=M_A+M_D$, $G$ is the gravitational constant, and $a$ is the 
semi-major axis of the system. The spin angular momenta of the accretor 
and donor are $J_{spin,A} = k_AM_AR_A^2\Omega_A$ and $J_{spin,D} = 
k_DM_DR_D^2\Omega_D$, respectively, where $k_A$ and $k_D$ are 
dimensionless constants depending upon the internal structure of the 
accretor and donor, respectively. It follows that:
\begin{multline}
\label{eq-3}
J_{tot} = \sqrt{\frac{Ga}{M}}M_AM_D \\
+ k_AM_AR_A^{2}\Omega_A + k_DM_DR_D^{2}\Omega_D.
\end{multline}

There are three effects that change the orbital angular momentum over 
time: mass transfer (MT), tides, and gravitational radiation (GR). 
Assuming each effect is indepedent of the others, we can then write the 
total change of the orbital angular momentum as the sum of the changes 
due to each of the above effects, with subscripts, as noted 
above, for each of the respective components:
\begin{equation}
\label{eq-4}
\dot J_{orb} = \dot J_{orb, MT} + \dot J_{orb, tides} + \dot J_{orb, 
GR}.
\end{equation}
To determine the total change in the angular momentum, then, we 
simply need to write the change due to each of the above effects.

The change in orbital angular momentum due to GR for a circular 
orbit is given by:
\begin{equation}
\label{eq-5}
\dot J_{orb, GR} = -\frac{32}{5}\frac{G^3}{c^5}\frac{M_AM_DM}{a^4}J_{orb}
\end{equation}
\citep[see for example,][]{Landau1975}.

Prior to the onset of mass transfer, the semi-major axis of the binary 
will have been shrinking due to the effects of GR as seen in 
equation~(\ref{eq-5}). During this time, tidal coupling will act to 
circularize the binary as well as synchronize the spins of the stars 
with the orbit. At the onset of mass 
transfer, we assume the spins and orbit to be synchronized. As mass 
transfer begins, angular momentum will be exchanged between the spins of 
the component stars and the orbit (see Paper I).  Any resulting 
asynchronization between the spins and the orbit leads to tidal coupling, the strength of which will determine how much of the spin angular 
momentum of the accretor is returned to the orbit.  This will ultimately affect the stability of the mass 
transfer process.

As in Gokhale2007, we model the change in the binary 
orbital angular momentum due to tides by:
\begin{equation}
\label{eq-6}
\dot J_{orb, tides} = \frac{k_AM_AR_A^{2}}{\tau_A}\omega_A + 
\frac{k_DM_DR_D^{2}}{\tau_D}\omega_D.
\end{equation}
The first term on the right-hand side of this equation represents the 
torque due to dissipative coupling upon the accretor, and the second 
term the torque upon the donor. These torques are parameterized in terms 
of the synchronization time-scales of the accretor, $\tau_A$, and donor, 
$\tau_D$ and are linearly proportional to the difference between the component and orbit spin, $\\omega_i = \Omega_{i} - \Omega_{orb}$, with 
$i\in\{A,D\}$ for the accretor and donor, respectively. Here, 
$\Omega_{orb}$ is the angular velocity of the circular orbit:
\begin{equation}
\label{eq-omega}
\Omega_{orb} = \sqrt{\frac{GM}{a^3}}
\end{equation}
and $k_iM_iR_i^2$ are the moments of inertia of each 
component, with $k_i$ being the inertial constant.

The values of $\tau_A$ and $\tau_D$ determine the strength of tidal 
coupling relative to MT and GR. As in Gokhale2007, we examine both the 
cases of $\tau_A = \tau_D = 10^{15}$ years (weak tidal coupling) and 
$\tau_A = \tau_D$ = 10 years (strong tidal coupling). The 
tidal-synchronization timescales are discussed further in 
\ref{sec:sync}.

To determine $\dot{J}_{orb,MT}$, we follow Paper I, which uses the 
ballistic mass transfer calculations of \citet{Sepinsky2010} to 
determine the instantaneous effect of mass transfer on DWD systems. This 
method uses a fully self-consistent, conservative, ballistic model of 
the transferred mass to determine the orbital parameters of the system 
after a single mass-transfer event.  As seen in that paper, provided the 
mass transferred is small compared to the total mass of the binary 
system, the change in the orbital parameters per unit mass that results 
is directly proportional to the mass transfer rate:
\begin{equation}
\dot{J}_{orb,MT}=\frac{\Delta J_{orb,b}}{M_P}\dot{M}_D
\end{equation}
where $\Delta J_{orb,b}$ is the change in the orbital angular momentum 
for the DWD as calculated by the above ballistic model for a single mass 
transfer event ejecting a particle of mass $M_P$. In this method, changes in $J_{orb,MT}$ (and changes in all orbital 
parameters) are calculated at each time-step by integrating the 
three-body system consisting of the two stars and the discrete particle 
representing the mass transfer stream. The change in the $J_{orb,MT}$
per unit mass transferred is independent of the mass of the ejected 
particle as long as $M_P << M_D, M_A$.  For the calculations here, we 
use $M_P = 10^{-8} M_{\odot}$. The rate of 
change of $J_{orb,MT}$ is then determined by multiplying by the current
mass-transfer rate of our evolving DWD, $\dot M_D$. The calculation of $\dot{M_D}$ will be discussed in section~\ref{sec:mt}.

\subsection{Differential equations for long-term evolution}

Using the above rates of change for the orbital angular momentum, we 
are now ready to develop the equations for long-term evolution.  It 
follows from equation~(\ref{eq-2}) that
\begin{equation}
\label{eq-7}
\frac{\dot J_{orb}}{J_{orb}} = (1-q)\frac{\dot{M_D}}{M_D} + 
 \frac{1}{2}\frac{\dot{a}}{a}
\end{equation}
for conservative mass transfer ($\dot M = 0$).\footnote{Recall that we assume 
the orbit remains circular throughout.}  We let $q = M_D/M_A$.

\subsubsection{Evolution of the semi-major axis}
\label{sec:adot} 

As in section~\ref{sec:angmom} for the 
orbital angular momentum, we examine the changes of the semi-major axis 
due to mass transfer, tides, and GR. We assume that each effect is 
independent, and write the total change in the semi-major axis due to 
each of the above effects, respectively, as:
\begin{equation}
\label{eq-10}
\dot{a} = \dot{a}_{MT} + \dot{a}_{tides} + \dot{a}_{GR}.
\end{equation}

To calculate $\dot{a}_{MT}$, we utilize the model for mass transfer 
developed in \citet{Sepinsky2010}. The differences between this model 
and models used in other analyses will be discussed in the section 
\ref{sec:diff}. Using this method, we calculate the time rate of change 
of $a$ due to mass transfer as
\begin{equation}
\label{eq-11}
\dot{a}_{MT} = \frac{\Delta a_b}{M_P}\dot M_D.
\end{equation}
Here, $\Delta a_b$ is the change in the semi-major axis for the DWD as 
calculated by the above ballistic model of \citet{Sepinsky2010} as described in section~\ref{sec:angmom} for the orbital angular momentum.

Next we calculate $\dot a_{tides}$ and $\dot a_{GR}$. We cannot use the 
standard tidal prescription for changes in the semi-major axis following 
\citet{Hut1981} because, following Marsh2004 and Gokhale2007, we used a 
different metric for changing the spin angular momentum due to tides 
(equation [\ref{eq-6}]). Instead, we must develop $\dot a_{tides}$ from 
that angular momentum change. Holding the donor mass constant ($\dot M_D = 0$) for the final two terms of 
equation~(\ref{eq-10}) we can combine equation~(\ref{eq-4}) with 
equation~(\ref{eq-7}) to obtain:
\begin{equation}
 \label{eq-9}
 \frac{\dot J_{orb,tides}}{J_{orb}} + \frac{\dot J_{orb,GR}}{J_{orb}} = 
 \frac{1}{2a} \dot a_{tides, GR}
 \end{equation}
where $\dot a_{tides, GR}$ is the total change of the 
semi-major axis due to the combined effects of tides and GR.
Assuming as before that changes to the semi-major axis due to tides and 
changes due to GR are independent we can re-write the above as:
\begin{equation}
\label{eq-12}
\frac{\dot J_{orb,tides}}{J_{orb}} + \frac{\dot J_{orb,GR}}{J_{orb}} = \frac{1}{2a} \dot a_{tides} + \frac{1}{2a} \dot a_{GR}.
\end{equation}
Using equations~(\ref{eq-2}), (\ref{eq-5}), and (\ref{eq-6}), it follows that:
\begin{multline}
 \label{eq-15}
 \dot a_{tides} = 2a\frac{1}{M_AM_D}\sqrt{\frac{M}{Ga}} \\
\times \left( \frac{k_AM_AR_A^{2}}{\tau_A}\omega_A + \frac{k_DM_DR_D^{2}}{\tau_D}\omega_D \right)
 \end{multline}
and
\begin{equation}
 \label{eq-16}
 \dot a_{GR} = - \frac{64}{5}\frac{G^3}{c^5}\frac{M_AM_DM}{a^3}
 \end{equation}

For our purposes, we rewrite the two tidal components in terms of the 
rotation rates of the accretor and the donor relative to the orbital 
angular velocity, which we define as $f_A$ and $f_D$, respectively. The 
rotation rates can be written in terms of the rotational angular 
velocities of the stars, $\Omega_i$ and the angular velocity of the 
circular orbit, 
$\Omega_{orb}$:
\begin{equation}
 \label{eq-13}
 f_i -1 = \frac{\Omega_i - \Omega_{orb}}{\Omega_{orb}} = \frac{\omega_i}{\Omega_{orb}}
 \end{equation}

Using equations~(\ref{eq-omega}) and (\ref{eq-13}), we  can rewrite equation~(\ref{eq-15}) as
\begin{equation}
\label{eq-14}
\dot a_{tides} = \frac{2M}{aM_AM_D} (\alpha_A +  \alpha_D)
\end{equation}
where
\begin{equation}
\alpha_A = \frac {k_AM_AR_A^2}{\tau_A}(f_A-1)
\end{equation}
and
\begin{equation}
\label{eq-alpha}
\alpha_D = \frac {k_DM_DR_D^2}{\tau_D}(f_D-1).
\end{equation}

Finally, we can insert equations~(\ref{eq-11}), (\ref{eq-16}), and 
(\ref{eq-14}) into equation~(\ref{eq-10}) to obtain the equation for 
 the evolution of the semi-major axis with time:
\begin{multline}
\label{eq-a}
\dot a = \frac{\Delta a_b}{M_P}\dot M_D
+ \frac{2M}{aM_AM_D} (\alpha_A + \alpha_D) \\
- \frac{64}{5}\frac{G^3}{c^5}\frac{M_AM_DM}{a^3}.
\end{multline}

\subsubsection{Evolution of the component rotation rates}

Next, we find the equations for the evolution of the component spins, 
$f_A$ and $f_D$. Like the changes to the semi-major axis 
(equation~[\ref{eq-10}]), the changes in $f_A$ and $f_D$ can be 
separated into three components:
\begin{equation}
\label{eq-20}
\dot{f}_i = \dot{f}_{i, MT} + \dot{f}_{i, tides} + \dot{f}_{i, GR}
\end{equation}
with $i\in\{A,D\}$.

Analogous to $\dot a_{MT}$ in equation~(\ref{eq-11}), the change in 
$f_i$ due to mass transfer, $\dot{f}_{i, MT}$ can be written as:
\begin{equation}
\label{eq-21}
\dot f_{i, MT} = \frac{\Delta f_{b,i}}{M_P}\dot M_D.
\end{equation}
Here $\Delta f_{b,i}$ is the change in the rotation rates of star $i$ 
resulting from a single mass transfer event in the formulation of 
\citet{Sepinsky2010} as described in section~\ref{sec:angmom}.

As in equation~(\ref{eq-3}), the spin angular momentum of each component can 
be written as
\begin{equation}
\label{eq-22}
J_{spin, i} = k_iM_iR_i^2f_i\Omega_{orb}.
\end{equation}
where we have substituted $f_i$ from equation~(\ref{eq-13}).

Since we have already determined the change in $f_i$ due to mass 
transfer (equation~[\ref{eq-21}]), we can determine $\dot{f}_{i,tides}$ 
and $\dot{f}_{i,GR}$ by differentiating equation~(\ref{eq-22}) with the 
mass held constant:
\begin{multline}
\label{eq-23}
\left.\dot{J}_{spin,i} \right|_{M_i} = k_iM_iR_i^2\Omega_{orb}(\dot{f}_{i,tides} + 
\dot{f}_{i,GR}) \\
-\frac{3}{2}k_iM_iR_i^2\Omega_{orb} f_i \frac{1}{a} (\dot{a}_{tides} + 
\dot{a}_{GR})
\end{multline}
where the second term arises due to the dependence of $\Omega_{orb}$ on the semi-major axis (equation~[\ref{eq-omega}]). We note that, because we are holding the mass constant, the second term depends only upon changes due to tides and GR, and not changes due to mass transfer. Changes to $f_i$ due to the effect of mass transfer on the semi-major axis are fully accounted for by equation~(\ref{eq-21}).

Since we do not include any GR effects on the spin angular 
momentum of the components, conservation of angular momentum dictates 
that any changes in the spin angular momentum of a component must be 
equal and opposite to the changes in the orbital angular momentum of the 
system due to tides acting on that component. Combining 
equations~(\ref{eq-6}), (\ref{eq-omega}), and (\ref{eq-13}), we have:
\begin{equation}
\label{eq-25}
\left.\dot{J}_{spin,i}\right|_{M_i} = 
-\frac{k_iM_iR^2_i}{\tau_i}\Omega_{orb}(f_i-1).
\end{equation}

By combining equations~(\ref{eq-23}) and (\ref{eq-25}) and rearranging, 
we can write $\dot f_{GR, i} + \dot f_{tides, i}$ as:
\begin{multline}
 \label{eq-26}
 \dot f_{GR, i} + \dot f_{tides, i} = -\frac{(f_i-1)}{\tau_i} + \\
 \frac{3}{2}\frac{f_i}{a}\left[-\beta + \frac{2}{a}\frac{M}{M_AM_D}(\alpha_A + \alpha_D) \right]
 \end{multline}
where we have used equations~(\ref{eq-16}), (\ref{eq-14})--(\ref{eq-alpha}), and let:
\begin{equation}
 \beta = \frac{64}{5}\frac{G^3}{c^5}\frac{M_AM_DM}{a^3}.
 \end{equation}

Following the form of equation~(\ref{eq-20}), we can combine 
equations~(\ref{eq-21}) and (\ref{eq-26}) to write the equations for the 
evolution of $f_A$ and $f_D$:
\begin{eqnarray}
\label{eq-27}
\dot f_A = \frac{\Delta f_{b,A}}{M_P}\dot M_D - \frac{f_A-1}{\tau_A} + \\ \nonumber
\frac{3}{2}\frac{f_A}{a}\left[-\beta + \frac{2}{a}\frac{M}{M_AM_D}(\alpha_A+\alpha_D) \right]
\end{eqnarray}

\begin{eqnarray}
\label{eq-28}
\dot f_D = \frac{\Delta f_{b,D}}{M_P}\dot M_D - \frac{f_D-1}{\tau_D} + \\ \nonumber
\frac{3}{2}\frac{f_D}{a}\left[-\beta + 
\frac{2}{a}\frac{M}{M_AM_D}(\alpha_A+\alpha_D)\right].
\end{eqnarray}

\subsubsection{Evolution of the eccentricity}

We assume that the eccentricity of these systems is zero prior to the 
onset of direct-impact accretion. During the three-body integration of a 
single mass-transfer event, it is possible for the binary to develop a 
small eccentricity. However, for simplicity and in accordance 
with Marsh2004 and Gokhale2007, we force the eccentricity to remain zero 
throughout. Because the system begins in a circular orbit, and due to 
the action of tidal forces and gravitational wave emission which both 
act to circularize the orbit, it is unlikely for any significant 
eccentricity to develop. In order to keep this orbital angular momentum 
in the orbit, we manually set our new semi-major axis to:
\begin{equation}
\label{eq-18}
a = a_I (1-e^2)
\end{equation}
where $e$ is the eccentricity gained by the system during mass transfer and $a_I$ is the semi-major axis of the eccentric orbit. Using 
this modification, we force:
\begin{equation}
\label{eq-19}
\dot e = 0,
\end{equation}
and still conserve orbital angular momentum. A more thorough analysis in 
which we allow eccentricity to vary throughout the entire calculation 
will be presented in a forthcoming paper.

\subsubsection{Mass transfer rates}

Finally, the changes in the masses of the accretor and donor are given, 
respectively, by:
\begin{equation}
\label{eq-29}
\frac{dM_A}{dt} = -\dot M_D
\end{equation}
\begin{equation}
\label{eq-30}
\frac{dM_D}{dt} = \dot M_D
\end{equation}
The mass loss rate of the donor, $\dot M_D$ is obtained as described in 
section~\ref{sec:mt}.

\bigskip

Together, the set of equations~(\ref{eq-a}), (\ref{eq-27}), (\ref{eq-28}), (\ref{eq-19}), (\ref{eq-29}), (\ref{eq-30}) can be integrated in time to calculate the evolution of the system.

\subsection{Differences from previous analyses}
\label{sec:diff} 

There are several differences between our treatment of mass transfer and 
those of Marsh2004 and Gokhale2007. In order to calculate the angular 
momentum exchange during mass transfer, the studies of Marsh2004 and 
Gokhale2007 both utilize a numerical prescription based on 
\citet{Verbunt1988}. In that formulation, it is assumed that the angular 
momentum transferred from the orbit to the spin of the accretor is 
exactly equal to the angular momentum of the ballistic particle in a 
circular orbit around the donor at its average 
radius during its motion from donor to accretor. Where the analysis of 
Marsh2004 keeps the spin of the donor fixed, Gokale2007 does allow the 
spin of the donor to vary, and notes that in doing so, the number of 
stable systems increases. However, the allowance of variation in 
donor spin is not done self-consistently. By introducing the spin 
angular momentum of the donor, there are now three separate 
sources/sinks of angular momentum: the spin of the donor, the spin of 
the accretor, and the orbit.  Paper I showed that angular momentum is 
transferred between each during mass transfer, and that the fraction of 
angular momentum transferred between each is strongly dependent on the 
ballistic trajectory of the transferred mass, and hence the system 
properties.  In this paper, we apply the ballistic calculations of Paper 
I to lift the dependence on the \citet{Verbunt1988} approximations to 
more accurately determine the flow of angular momentum between the 
component spins and the orbit.

Since orbital angular momentum changes are directly linked to changes in 
the mass ratio and semi-major axis, much can be learned from analyses 
such as that of Gokhale2007 and Marsh2004. However, a more complete 
analysis demands a thorough consideration of the spin angular momenta of 
the stars as well due to their indirect effect upon the system 
properties through tidal coupling. This is accomplished via the 
ballistic calculations described above, which evaluate the three-body 
problem throughout the evolution of the system to determine the precise 
effect mass transfer has on the evolution of the system. These 
calculations take into account not only the immediate feedback on the 
orbit and spin of the accretor during ejection, but also (1) the 
gravitational effect on the orbit during the mass transfer process, (2) 
a calculation of the precise moment when the particle impacts the 
surface of the accretor, (3) the instantaneous properties of both the 
accretor and particle at impact, including the angle of impact, and (4) 
correctly divides the momentum of the impacting particle between linear 
and angular momentum based upon the angle of impact and the rotation 
rate of the accretor. We can then examine the angular momentum exchange 
between all components, the spins of both stars and the orbit, and do so 
in a way that allows the spins of both stars to vary self-consistently.

\subsection{Calculation of mass-transfer rate}
\label{sec:mt} 

As in Marsh2004,  we define the overfill of the Roche lobe as:
\begin{equation}
\label{eq-31}
\Delta = R_D-R_L
\end{equation}
where $R_D$ is the radius of the donor and $R_L$ is the radius of the 
donor's Roche lobe. The way in which the mass-transfer rate varies with 
$\Delta$ has been investigated in many analyses (see, for example, 
\citet{Paczynski1972,Webbink1977,Savonije1978}). In accordance with 
Marsh2004, we approximate the mass transfer as adiabatic 
\citep{Webbink1984}. In the adiabatic regime the mass transfer 
rate is given by:
\begin{eqnarray}
\label{eq-32}
\dot{M_D} = \frac{8\pi^3}{9} \left(\frac{5Gm_e}{h^2}\right)^{3/2} 
 (\mu_e m_n)^{5/2} \times \nonumber \\
\frac{1}{P_{orb}} \left(\frac{3\mu M_D}{5r_LR_2}\right)^{3/2} 
 \frac{1}{\sqrt{a_2(a_2-1)}}\Delta^3
\end{eqnarray}
for $\Delta > 0$ and zero for $\Delta < 0$ (using results from 
\citet{Webbink1984,Chandrasekhar1967,Webbink1977}). Here, $m_e$ is the 
mass of an electron, $m_n$ is the mass of a nucleon, $\mu_e$ is the mean 
number of nucleons per free electron in the outer layers of the donor 
(assumed here to be two), $P_{orb}$ is the orbital period, $r_L = R_L/a$, and $\mu$ 
and $a_2$ are given by the following:
\begin{equation}
\label{eq-33}
\mu = \frac{M_D}{M_A+M_D}
\end{equation}
\begin{equation}
\label{eq-34}
a_2 = \frac{\mu}{x_{L1}^3} + \frac{1-\mu}{(1-x_{L1})^3}
\end{equation}
where $x_{L1}$ is the distance from the center of the donor to the inner 
Lagrangian point of the donor, in units of the semi-major axis 
\citep{Webbink1977}.

\subsection{Calculation of Roche lobe}

In the case of Marsh2004, where the rotational velocity of the donor is 
fixed to the orbital velocity and where the orbit is circular 
throughout, the shape and volume of the Roche lobe depends only upon the 
mass ratio of the system. In that case, the approximation from 
\citet{Eggleton1983},
\begin{equation}
\label{eq-35}
R_{L,{\rm Egg}}  = a \frac{0.49 q^{2/3}}{0.6 q^{2/3}+ln(1+q^{1/3})},
\end{equation}
is sufficient. However, for eccentric and/or non-synchronous binaries, 
the shape and volume of the roche lobe also depend upon the 
eccentricity, the true anomaly, $\nu$, and the rotation rate, which can 
be expressed by the parameter:
\begin{equation}
\label{eq-36}
\mathcal{A}_i(e,f,\nu) = \frac{f_i^2(1+e)^4}{(1+e \cos \nu)^3}
\end{equation}
see \citet{Sepinsky2007}. Other analyses have considered the dependence of the Roche lobe on asynchronicity. In the case of a circular orbit, the Roche lobes described in \citet{Sepinsky2007} using the Roche-lobe parameter $\mathcal{A}_i$ directly reflect the equipotential surfaces described in \citet{Plavec1958} and \citet{Kruszewski1963}. We note that this 
asynchronicity parameter ${\cal A}_i$ depends on the rotation rate of 
the component star in question, which means the size of the Roche lobe 
can be different for each object in the system.

For our calculations, we perform a Monte Carlo integration to calculate 
the volume-equivalent Roche lobe radius (as outlined in 
\citet{Sepinsky2007}) over a two-dimensional grid in $\mathcal{A}$-$q$ 
parameter-space. We then use a bilinear interpolation function to create 
a continuous function for the Roche lobe from the two-dimensional grid. 
The interpolated function is accurate to 0.01 percent or better in 
comparison to the Monte Carlo integration at equivalent points.

In section~\ref{sec:rlasynch}, we examine the dependence of the results 
on the size of the Roche lobe.  There, we compare the evolution of the 
DWD systems using the Eggleton function for the roche lobe, $R_{L,{\rm 
Egg}}$, to the evolution including the asynchronicity parameter which we 
name $R_{L,asynch}$.

\subsection{Synchronization time-scale}
\label{sec:sync} 

As the spin angular momentum of the accretor increases due to impact 
from the mass transfer stream, tidal coupling will redistribute some of 
this angular momentum back into the orbit. The strength of this tidal 
coupling is dependent upon the synchronization time-scales, $\tau_A$ and 
$\tau_D$, as seen in equation~(\ref{eq-6}). \citet{Campbell1984} calculated this timescale as:
\begin{equation}
\label{eq-37}
\tau_C = 1.3 \times 10^{7} \left(\frac{M_A}{M_D}\right)^2 \left(\frac{a}{R_A}\right)^6 \left[\frac{M_A/M_{\odot}}{L_A/L_{\odot}}\right]^{5/7} yr
\end{equation}

Following Marsh2004, we retain the scaling with mass ratio and 
orbital separation, but, because of the uncertainty of the overall 
magnitude of the time-scale, we allow the magnitude to vary. We define 
the overall magnitude by the time-scale at the moment of first contact 
and assume, as in Marsh2004 and Gokhale2007, that
\begin{equation}
\label{eq-38}
\tau_A(t) =  C_A \left(\frac{M_A}{M_D}\right)^2 \left(\frac{a}{R_A}\right)^6
\end{equation}
and
\begin{equation}
\label{eq-39}
\tau_D (t) = C_D \left(\frac{M_D}{M_A}\right)^2 \left(\frac{a}{R_D}\right)^6.
\end{equation}
where $C_A$ and $C_D$ are constants defined such that $\tau_A(0) 
=\tau_{A,I}$ and $\tau_D(0)=\tau_{D,I}$. Here, $\tau_{A,I}$ and 
$\tau_{D,I}$ are the initial synchronization time-scales. As in Gokhale2007, we explore two different values for the synchronization 
time-scale at contact: $10^{15}$ years (very weak tidal coupling) and 10 years (very strong tidal coupling). Much work on tidal synchronization in DWDs has been done (see, for example, \citet{Valsecchi2013, Fuller2014, Burkart2014}). These analyses have shown that shorter synchronization time-scales may be a better approximation for systems with short orbital periods, as we consider in this analysis, which may lend credence to our 10 year synchronization time-scale.
For simplicity, we use the same initial time-scale for the donor and 
accretor ($\tau_{A,I}=\tau_{D,I}$) and refer to the initial synchronization time-scale for both stars as simply $\tau$.

\bigskip

We have now determined the set of differential equations which governs 
the evolution of the DWDs undergoing direct-impact accretion, including 
the methods for calculating the Roche lobe size, the mass transfer rate, 
and the synchronization timescales which determine the strength of tidal 
coupling. We now have all the steps in place to solve for the long-term 
evolution of the systems.

\section{Numerical Solutions}

We integrate equations~(\ref{eq-a}), (\ref{eq-27}), (\ref{eq-28}), 
(\ref{eq-19}), (\ref{eq-29}), (\ref{eq-30}) using an 8th order Runge 
Kutta ordinary differential equation solver \citep{Galassi2006}. 
Excluding the losses due to gravitational radiation, the total energy 
and momentum of the system are conserved throughout the integration over 
the entire parameter space to one part in $10^{-3}$ or better. At the 
end of the evolution, we calculate $J_{conserve} = J_{total} - J_{total,I} + J_{total, GR}$ where $J_{total,I}$ is the initial total angular momentum and $J_{total, GR}$ is the total angular momentum 
lost to gravitational radiation. If angular momentum is perfectly 
conserved, we expect $J_{conserve}$ to be equal to zero. 

Figure 1 shows 
the final value of $J_{conserve}$ for both synchronization time-scales 
at contact over the entire parameter space of interest for solutions 
calculated using both the Eggleton approximation (left), $R_{L,{\rm Egg}}$, and 
$R_{L,asynch}$ (right) for the calculation of the Roche lobe. The 
colors correspond to different values of $J_{conserve}$ as described in 
the caption.  In general, systems with a smaller total mass conserve 
total angular momentum better than systems with a higher mass.  For many 
of the systems in the lower right, this is due to the fact that they reach a stable configuration (disc accretion) 
early in the integration.  Therefore, the integration needs to run for 
only a few orbits.  Systems where $J_{conserve}$ is larger tend to be 
systems where the numerical integration runs for many orbits, allowing 
systematic errors in the numerical integration to accumulate.  Even so, 
all systems presented in this paper conserve total angular momentum to 
better than 1\% throughout the entire evolution.

We integrate over a period of 1 Gyr. As in Marsh2004, if 
$\dot{M_D}$ exceeds 0.01 $M_{\odot}/$yr at any point during the 
integration, the integration stops and the system is discarded as 
unstable.

As a system evolves, it is possible to pass back and forth through 
phases of mass transfer and phases of no mass transfer as the semi-major 
axis and the two masses change. If the semi-major axis increases enough for mass transfer to stop altogether, the integration proceeds (with $\dot M_D = 0$) until the action of gravitational radiation shrinks the orbit sufficiently for mass transfer to resume.

As in Paper I, in this 
work we are only interested in direct-impact mass transfer where the particle 
impacts the surface of the accretor within one orbital period. In this case, the evolution of the orbital parameters is determined by the differential equations presented in section 2. If the particle does not accrete within one orbital 
period, it is likely that the accretion stream will eventually intersect with 
itself, ultimately leading to the formation of an accretion disc \citep{Sepinsky2010}. 

\begin{figure} [t!]
\label{fig-conserve}
\plotone{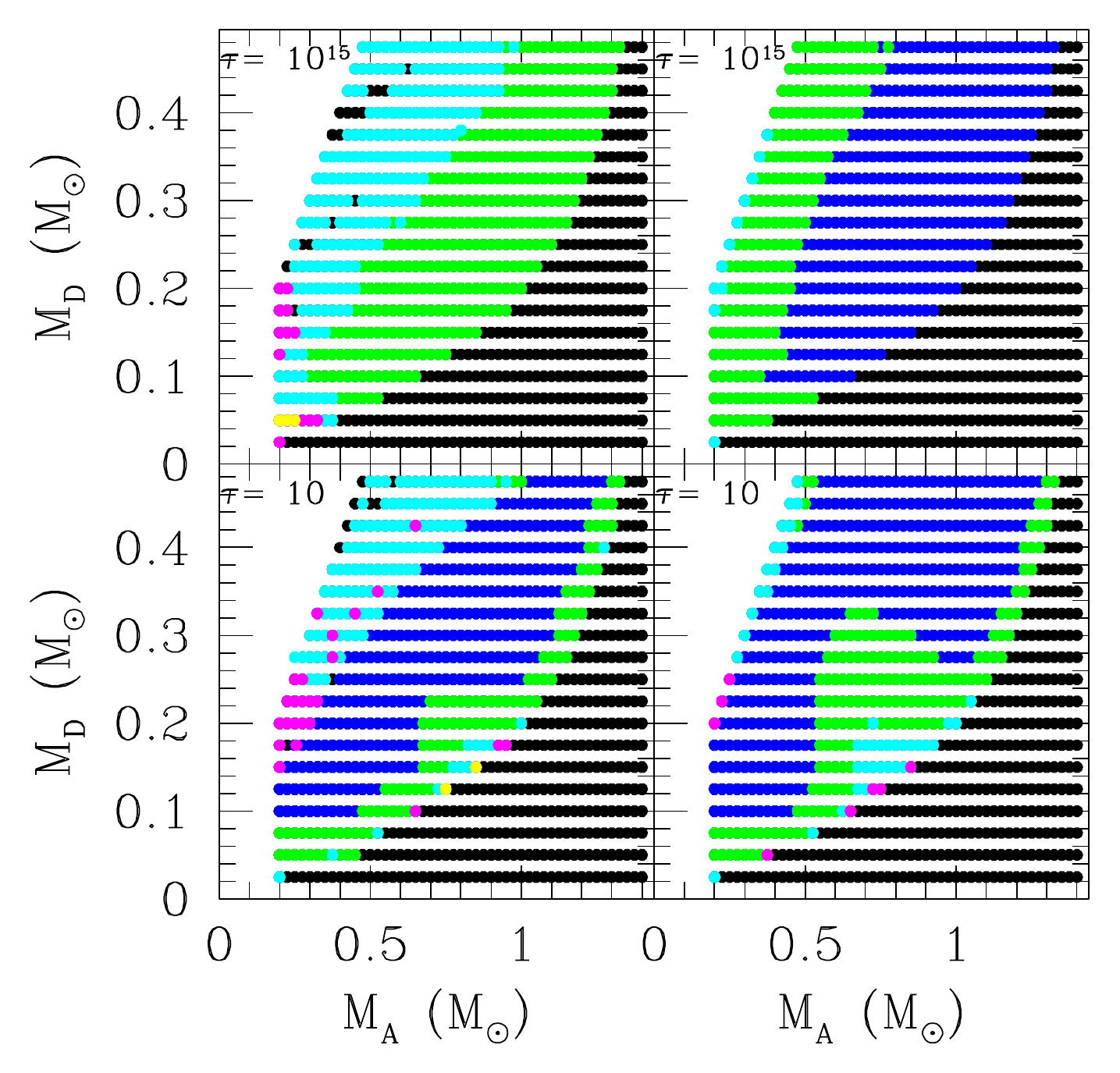}

\caption{ \footnotesize Angular momentum conservation over entire parameter space for the two synchronization time-scales. The left-hand panels show the solutions obtained using $R_{L,{\rm Egg}}$ for the calculation of the Roche lobe and the right-hand panel shows the solutions using $R_{L,asynch}$. Calculated here is $J_{conserve} = J_{total} - J_{total, I} + J_{total, GR}$ at the end of the integration. Blue systems conserve angular momentum to an order of $10^{-3}$; green systems to an order of $10^{-4}$; cyan systems to an order of $10^{-5}$; magenta systems to an order of $10^{-6}$; yellow systems to an order of $10^{-7}$; and black systems to an order of $10^{-8}$ or better.}
\label{fig:Figure1} 
\end{figure}

Compared to direct-impact accretion, disc accretion is known to provide a much 
more efficient mechanism for redistributing spin angular momentum in the 
accretor back into the orbit through tidal coupling \citep{Frank2002}. As a result, it is likely 
that once a system enters a phase of disc accretion it will remain in 
this phase or perhaps even become detached as the orbital separation 
continues to grow and the mass ratio decreases due to continued mass 
transfer. If at any point during our numerical integrations a disc is formed under these circumstances, the integration stops and the 
system is assumed to be stable throughout its lifetime. A more thorough 
analysis which continues to track the evolution through potential disc 
phases will be performed in a future paper. This issue is 
discussed further in section 4.

\subsection{Maximum spin-rate of the accretor}

For weak tidal coupling, it is possible for the accretor to be spun up to its breakup rate ($\Omega_k = \sqrt{GM_A/R_A^3}$). We track $\Omega_A$ throughout our calculations and note that only one system ever reaches this maximum spin for the accretor. This system, which has an initial donor mass of $0.1\,M_{\odot}$ and an initial accretor mass of $0.275\,M_{\odot}$, is marked by a yellow dot in Figure 2. The accretor spin-rates in all other systems stay below the breakup rate throughout their evolution.

\subsection{Super-Eddington accretion}

As noted by Marsh2004, there are likely ranges of parameter space where 
systems, despite being stable, experience super-Eddington accretion at 
some point during their evolution. It is expected that a possible 
ultimate consequence of sustained super-Eddington accretion is a merger 
\citep{Han1999,Nelemens2001,Marsh2004}, the same result as a dynamically 
unstable system, as they eventually can reach very high mass-transfer 
rates.

As in Marsh2004, we calculate the Eddington accretion rate using a modified form of the calculation used by \citet{Han1999}:
\begin{equation}
\label{eq-43}
\dot M_{Edd} = \frac{8\pi G m_p c M_A}{\sigma_T (\phi_{L1} - \phi_a -\frac{1}{2}\mathbf{v_i}^2 + \frac{1}{2}(\mathbf{v_i - v_{\omega}})^2)}
\end{equation}
where $\sigma_T$ is the Thomson cross-section of the electron, $m_p$ is the mass of a proton, $\mathbf{v_i}$ is the impact velocity of the accreted particle, and $\mathbf{v_{\omega}}$ is the spin-velocity of the accretor's surface at the point of impact, both measured in the co-rotating frame of reference.

In the Marsh2004 analysis, a set of impact velocities and locations were pre-computed and were interpolated during the calculations to calculate $\mathbf{v_i}$ and $\mathbf{v_{\omega}}$. Instead we are able to calculate  $\mathbf{v_i}$ and $\mathbf{v_{\omega}}$ explicitly as part of the three-body integration of the mass transfer event. We track the mass-transfer and flag systems that eventually exceed the maximum mass-transfer as unstable. 

\section{Results}

\begin{figure} [t!]
\plotone
{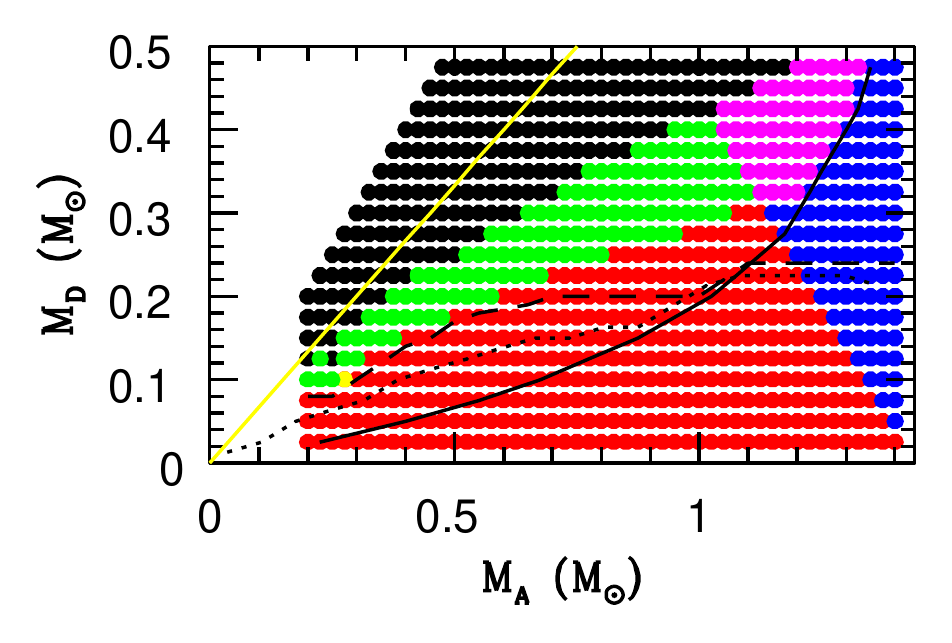}
\caption{\label{fig:1e15}  \footnotesize Results of evolution with initial tidal synchronization time-scale of $\tau = 10^{15}$ years using the Eggleton approximation for the Roche lobe. Red systems are stable throughout their entire evolution and have a total system mass which is sub-Chandrasekhar ( $< 1.44\,M_{\odot}$). Green systems have a total mass which is sub-Chandrasekhar but went through a phase of super-Eddington accretion at some point during the evolution. Blue systems are stable throughout and have a total system mass which is super-Chandrasekhar. Magenta systems have a total mass that is super-Chandrasekhar and undergo a period of super-Eddington accretion. Black systems are unstable, meaning the mass-transfer rate exceeded $0.01\,M_{\odot}/$yr. The solid yellow line illustrates the widely-used $q > 2/3$ instability boundary (see, for example, Marsh2004). The solid black line shows the boundary between disc and direct-impact accretion for initially synchronous and circular binaries. The dotted black line and dashed black line illustrate the boundary between stable systems and super-Eddington systems for the $\tau = 10^{15}$ year timescale in Marsh2004 and Gokhale2007, respectively (see Figure 5 in Marsh2004 and Figure 7 in Gokhale2007).}
\end{figure}

Using the techniques described above, we computed evolution over a grid 
in $M_A$,$M_D$ parameter space to determine the long-term stability of various systems. The grid was computed for two 
different tidal synchronization time-scales at contact: $10^{15}$ years and 10 
years.

\subsection{Evolution of systems using Eggleton Roche lobe}

We first present the results of our numerical integrations using the standard Eggleton approximation for the calculation of the Roche lobe (equation~[\ref{eq-35}]), which does not take into account asynchronicity between the donor and orbit. The Eggleton approximation was was used in the analyses of both Marsh2004 and Gokale2007.

Figure 2 shows the end result of systems with an initial synchronization 
time-scale of $\tau = 10^{15}$ years using $R_{L,{\rm Egg}}$. 
As mentioned in section 3, we assume that once a disc is reached, the 
system will remain stable for the remainder of its evolution, therefore 
we stop the integration once a disc is formed. With the exception of the 
unstable black systems (whose evolution stops when the mass transfer 
rate exceeds the maximum limit of $0.01~M_{\odot}/$yr), the evolution of 
all systems in this plot is stopped when the system reaches a phase of 
disc accretion. The time it takes to reach this phase 
of disc accretion varies from system to system.

\begin{figure} [b!]
\plotone
{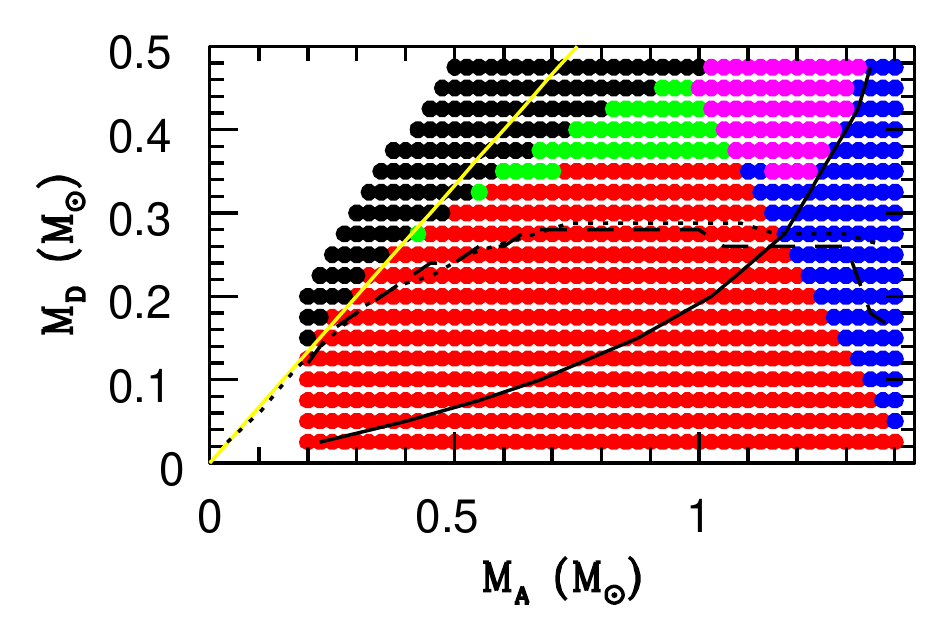}
\caption{ \footnotesize As Figure 2, but with $\tau = 10$ years.}
\end{figure}

The solid black line in Figure 2 shows the boundary between disc and 
direct-impact accretion for initially synchronous and circular binaries. 
All systems to the right of and below this line begin evolution in a 
disc phase. Therefore, these systems are assumed stable from the onset 
of evolution and do not evolve past $t = 0$. Since these systems do not evolve in time, the angular momentum is perfectly conserved, as shown by the black systems in the lower right of the plots in Figure 1. All systems to the left of and above this line begin evolution 
in a direct-impact phase. We will discuss how we handle disc accretion 
further in section 4.2.

Figure 3, analogous to Figure 2, shows  the end results of systems with initial
synchronization time-scales of $\tau = 10\,$years using $R_{L,{\rm Egg}}$ for the Roche lobe calculation.

The sub-Chandrasekhar, sub-Eddington systems shown in red in Figures 2 and 3 are expected to remain stable throughout their lifetimes and are therefore characterized as likely AM CVn progenitors. As observed from the dotted and dashed black lines in Figures 2 and 3, the region of parameter space occupied by such systems in Gokhale2007 (dashed line) increased the number of stable systems compared to Marsh2004 (dotted line); our analysis reveals a further increase in the extent of this region of parameter space. This observation holds true for both the $10^{15}$ year time-scale (Figure 2) and the 10 year time-scale (Figure 3).

Gokhale2007 noted that by allowing the spin of the donor to vary and by 
including the resulting effects of tidal coupling between the donor's 
spin and the orbit, the number of stable systems increased compared to 
the analysis of Marsh2004 (as observed by comparing the dotted versus 
dashed lines in both Figures 2 and 3). As described in section 2.4, the 
main difference between our analysis and that of Gokhale2007 is in the 
treatment of the way we handle angular momentum evolution during the 
direct impact accretion process. In observing the increase in the number 
of stable systems compared to the analysis of Gokhale2007, we conclude 
that by utilizing a mass transfer treatment that allows the rotation 
rates of both components to vary and self-consistently accounts for the 
exchange of angular momentum between the spins of the components and the 
orbit, we are able to increase the number of stable systems.

Figures 2 and 3, produced using $R_{L,{\rm Egg}}$ for the Roche lobe 
calculation, isolate the effect of our different treatment of the mass 
transfer process has upon the stability of systems in comparison with 
Marsh2004 and Gokhale2007. We will discuss the differences resulting 
from using $R_{L,asynch}$ in section 4.4.

Forming a boundary between the stable (red and blue) and unstable 
(black) systems are the green and magenta systems, which experience 
super-Eddington accretion at some point during the integration. Unlike 
the black systems, whose mass transfer exceeds the allowed limit of 
$0.01 M_{\odot}/$yr and are therefore categorized as unstable, these 
green/magenta super-Eddington systems are categorized as stable. 
However, it is likely that systemic mass loss is needed for a binary to survive 
phases of super-Eddington accretion. \citet{Han1999} argues that such 
mass loss may lead to a common envelope surrounding the binary, which 
will ultimately lead to a merger. In this case, systems which experience 
super-Eddington accretion would be considered unstable systems. The 
analyses of Marsh2004 and Gokhale2007 both note that super-Eddington 
systems are expected to be unstable. We make the same assumption but 
note that a more thorough treatment of the physics governing systems as 
they pass through any phases of super-Eddington accretion is necessary 
to state with certainty whether or not such systems are ultimately 
stable or unstable.

The green super-Eddington systems have a total mass which is 
sub-Chandrasekhar, and the magenta super-Eddington systems are 
super-Chandrasekhar. If we assume 
super-Eddington systems are ultimately unstable, 
the magenta systems can be categorized as possible Type Ia 
supernova progenitors.

As the tidal synchronization time-scale is reduced from $\tau= 10^{15}$ 
years in Figure 2 to $\tau = 10$ years in Figure 3, we see the parameter 
space occupied by stable systems grows (red). 
This is to be expected and in agreement with the analyses of Marsh2004 
and Gokhale2007. Stronger tidal coupling will allow more spin angular 
momentum from the accretor to be transferred back into the orbit, which increases the semi-major axis, causing the mass-transfer rate to decrease. This results in an
increase the stability of the systems in general.

\subsection{Disc accretion}

\begin{figure} [t!]
\plotone{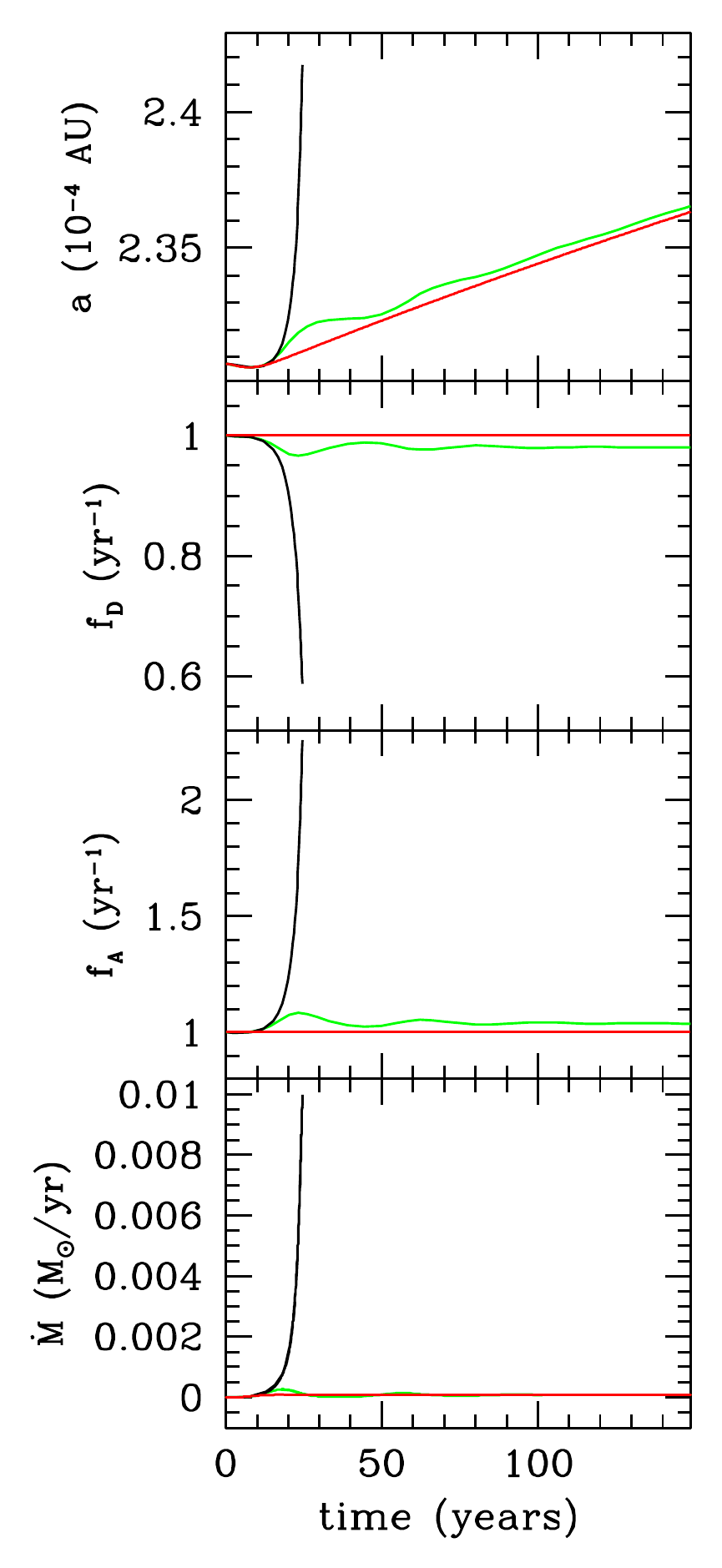}
\caption{\label{fig:Figure4} \footnotesize Various system properties for the first 150 years of evolution of a system with $M_{A} = 0.38\,M_{\odot}$ and $M_{D} = 0.8\,M_{\odot}$. Evolution using an initial synchronization time-scale of $\tau=10^{15}$ years is shown in black; this system is unstable (see Figure 3).  The same system with $\tau$ = 10 years is shown in green; this system is super-Eddington. $\tau = 0.1\,$ year for this system is shown in red; this system is stable.}
\end{figure}

In both Figures 2 and 3 we see a portion of stable, super-Chandrasekhar, 
sub-Eddington systems (blue) in the top right corners of our plots. In 
the analyses of both Marsh2004 and Gokhale2007, this portion of 
parameter space is occupied by super-Eddington systems, which would be 
magenta circles in our plots. Notice that the dotted and dashed lines of 
Marsh2004 and Gokhale2007 decrease as we move right across the plot, 
whereas our stability boundary is more extended. This is due to the fact 
that we do not follow the evolution of the systems past the development 
of disc accretion, while Marsh2004 and Gokhale2007 both continue to 
track the evolution of systems through phases of disc accretion.

Here we acknowledge the main limitation of our decision to assume that 
systems are stable once they reach a phase of disc accretion. The blue 
systems in the region of parameter space of interest here begin the 
integration in a disc phase, and are therefore labeled as stable. 
However, it is feasible that if these high donor-mass systems were 
allowed to evolve through the initial disc phase, they would eventually 
become super-Eddington and potentially unstable. In this case, 
this region of parameter space would more closely conform with the shape 
of Marsh2004's and Gokhale2007's analyses, both of which allow the 
systems to continue evolving through any phases of disc accretion. As we 
noted earlier, a more thorough analysis of this region of parameter 
space which includes treatment of disc accretion will be studied in a 
forthcoming paper.

\subsection{Comparison of system evolution with different tidal time-scales}

Figure 4 shows the evolution in time of a system with initial masses of 
$M_A = 0.8\,M_{\odot}$ and $M_D = 0.38\,M_{\odot}$ for the two different 
synchronization timescales at contact as well as a third much stronger 
tidal timescale, $\tau = 0.1$ year, for further illustration. These 
solutions are obtained using $R_{L,{\rm Egg}}$ to calculate the Roche 
lobe. For $\tau = 10^{15}\,$ years, the system is unstable, for $\tau = 
10\,$ years the system is stable, but passes through a phase of 
super-Eddington accretion, and for $\tau = 0.1\,$ year the system is 
stable and sub-Eddington.

For weak tidal coupling ($\tau = 10^{15}$ years), shown in black, mass 
transfer causes the rotation rates of the donor and accretor to rapidly 
decrease and increase, respectively. For the case of stronger tidal 
coupling ($\tau = 10$ years and $\tau = 0.1$ year; shown in green and 
red, respectively), tidal forces exist to redistribute angular momentum, 
working to keep the spins of the donor and accretor synchronous with the 
orbit ($f_i=1$).  

The rapid increase in the semi-major axis for $\tau = 10^{15}$ years is a 
direct result of the conservation of angular momentum: the angular 
momentum lost during the spin-down and mass loss of the donor is greater 
than the angular momentum gained by the accretor.  The net decrease in 
spin angular momentum corresponds to an increase in the orbital angular 
momentum, increasing the semi-major axis.  The mass transfer rate 
increases rapidly as the mass loss from the donor causes the radius to 
increase, even as the Roche lobe grows due to the increasing semi-major 
axis (see equations~[\ref{eq-31}] and [\ref{eq-32}]).

For the case of strong tidal coupling ($\tau=10, 0.1\,$year) we see a 
significant decrease in the mass transfer rate, with a correspondingly 
slower increase in the semi-major axis.  The rotation rates of the 
component white dwarfs remain closer to synchronous, and we do not see 
the rapid runaway that is evident in the case of weak tidal coupling.  

Finally, we note that oscillations in the orbital parameters are seen in 
the $\tau=10\,$ year case, which have been observed before by Gokhale2007.  
These oscillations are sensitive to small changes in the initial orbital 
parameters.  We discuss this phenomenon further in section~4.5.

\subsection{Evolution of systems including the effect of asynchronism on the Roche-lobe size}
\label{sec:rlasynch}

\begin{figure} [t!]
\plotone{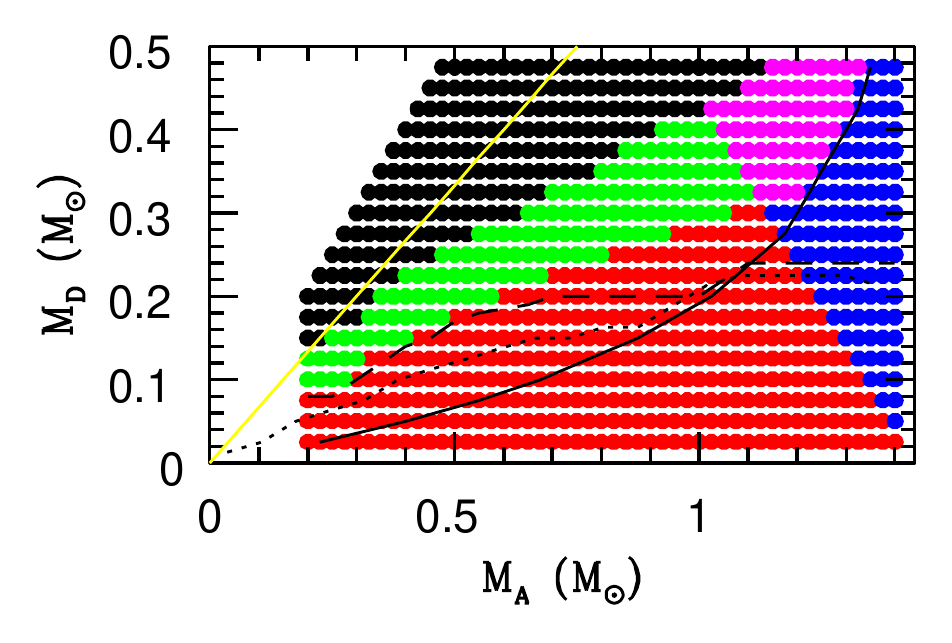}
\caption{\label{fig:01} \footnotesize As Figure 2, but using $R_{L,asynch}$ for the Roche lobe calculation.}
\end{figure}

Here, we show the results of our numerical 
integrations using $R_{L,asynch}$ to calculate the Roche-lobe size. As 
discussed in section 2.6, the Eggleton approximation is dependent upon 
only the mass ratio of the binary, whereas $R_{L,asynch}$ accounts for 
deviations from synchronism.

Figure 5 shows the end results of the systems with 
initial synchronization time-scales of $\tau = 10^{15}$ years 
and evolved using $R_{L,asynch}$. The color code 
remains the same as in Figures 2 and 3 and the dotted and dashed lines 
from Marsh2004 and Gokhale2007, respectively, for the $10^{15}$ 
year time-scale are included for reference. As can be seen, this plot is 
nearly identical to Figure 2, which was calculated using the Eggleton 
approximation.

Figure 6 shows the end results of the systems which begin with 
synchronization time-scales of $\tau = 10$ years.  By comparing with 
Figure 3, we observe that, unlike the $10^{15}$ year time-scale, $R_{L,asynch}$ has a 
significant effect upon the end-states of the systems for the 10 year 
time-scale. In particular, the number of unstable black systems is 
reduced substantially. The cause of this difference is explained in details below 
(section 4.5).

\begin{figure} [t!]
\plotone {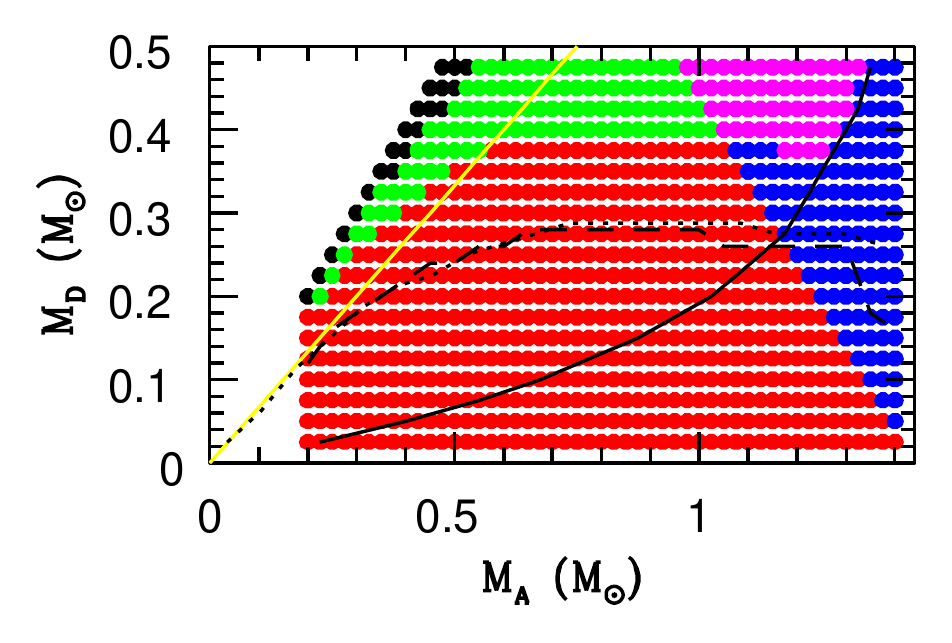}
\caption{ \footnotesize As Figure 3, but with $\tau = 10$ years and using $R_{L,asynch}$.}
\end{figure}

In Figure 6, we also observe a slight increase in the number 
of stable systems across the entire range of accretor masses compared to Figure 3. Additionally, there is a small group of stable red systems beyond the widely-used $q>2/3$ boundary for instability (for $q>2/3$, the donor radius will expand faster than the Roche lobe which suggests, simplistically, that such systems will become unstable). However, it is possible for systems with an initial mass ratio higher than 2/3 to survive, provided that the mass ratio is close to the boundary (see, for example, \citet{DSouza2006}). Our results, as shown in Figure 6, confirm this observation.

\subsubsection{Comparison of the evolution of systems using $R_{L,{\rm Egg}}$ and $R_{L,asynch}$ solutions for  $\tau = 10^{15}$ years}

\begin{figure} [t!]
\plotone{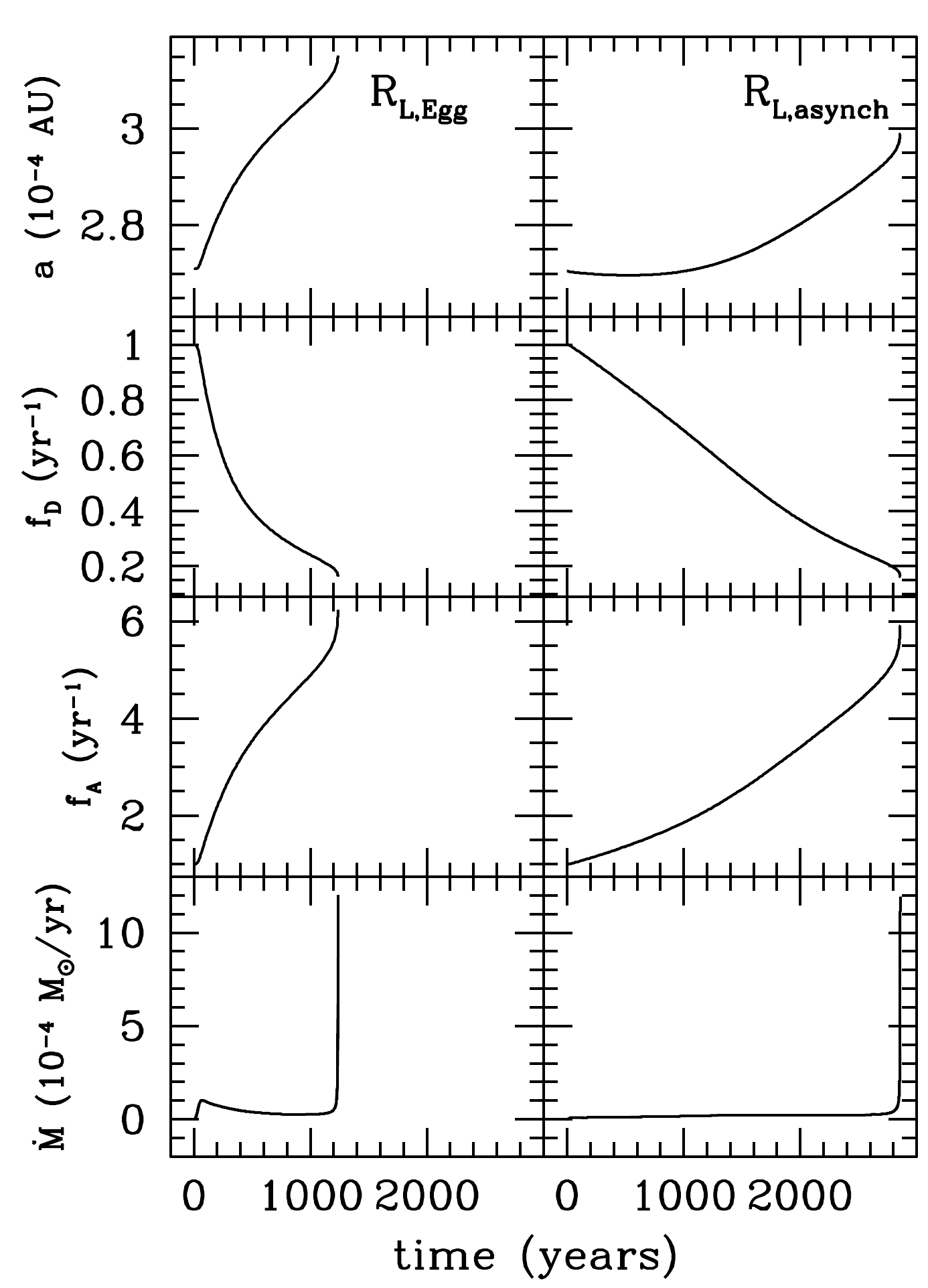}
\caption{ \footnotesize Evolution of orbital parameters for $\tau = 10^{15}$ years with initial masses of $M_D = 0.3~M_{\odot}$ and $M_A = 0.8~M_{\odot}$. Solutions obtained using the $R_{L,{\rm Egg}}$ calculation of the Roche lobe on left and the $R_{L,asynch}$ on the right.}
\end{figure}

Figure 7 compares the evolution of a system with an initial donor mass 
of 0.3 $M_{\odot}$ and an initial accretor mass of 0.8 $M_{\odot}$ for 
$\tau = 10^{15}$ years. On left 
we show the evolution of various orbital parameters obtained using the 
Eggleton approximation for the Roche lobe calculation. The right side 
mirrors the left but shows the evolution using $R_{L,asynch}$ instead of 
$R_{L,{\rm Egg}}$. As seen on the plots, the evolution of the orbital 
parameters happens on a much shorter time-scale for the $R_{L,{\rm 
Egg}}$ solution than for the $R_{L,asynch}$ solution. This is due to the fact that, as mass is 
transferred from the donor to the accretor, the rotation rate of the donor decreases relative to the 
orbit and the rotation rate of the accretor increases. As the donor spin ($f_D$) decreases, $\mathcal{A}$ (which is proportional to 
$f_D^2$; equation~\ref{eq-36}) also decreases. When 
using $R_{L,asynch}$, the size of the Roche lobe is inversely proportional to 
$\mathcal{A}$. So, as $\mathcal{A}$ decreases, the Roche lobe will 
increase, which ultimately reduces the mass transfer rate. However, in 
the $R_{L,{\rm Egg}}$ calculation, the Roche lobe is insensitive to 
changes in $f_D$, so the mass transfer rate is higher relative to that 
calculated in the solution using $R_{L,asynch}$. Since the mass transfer 
rate is higher in the Eggleton case, the orbital parameters change more 
quickly than in the $R_{L,asynch}$ case, as seen in Figure 7.

\subsubsection{Comparison of the evolution of systems using $R_{L,{\rm Egg}}$ and $R_{L,asynch}$ solutions for $\tau$ = 10 years}

\begin{figure} [t!]
\plotone{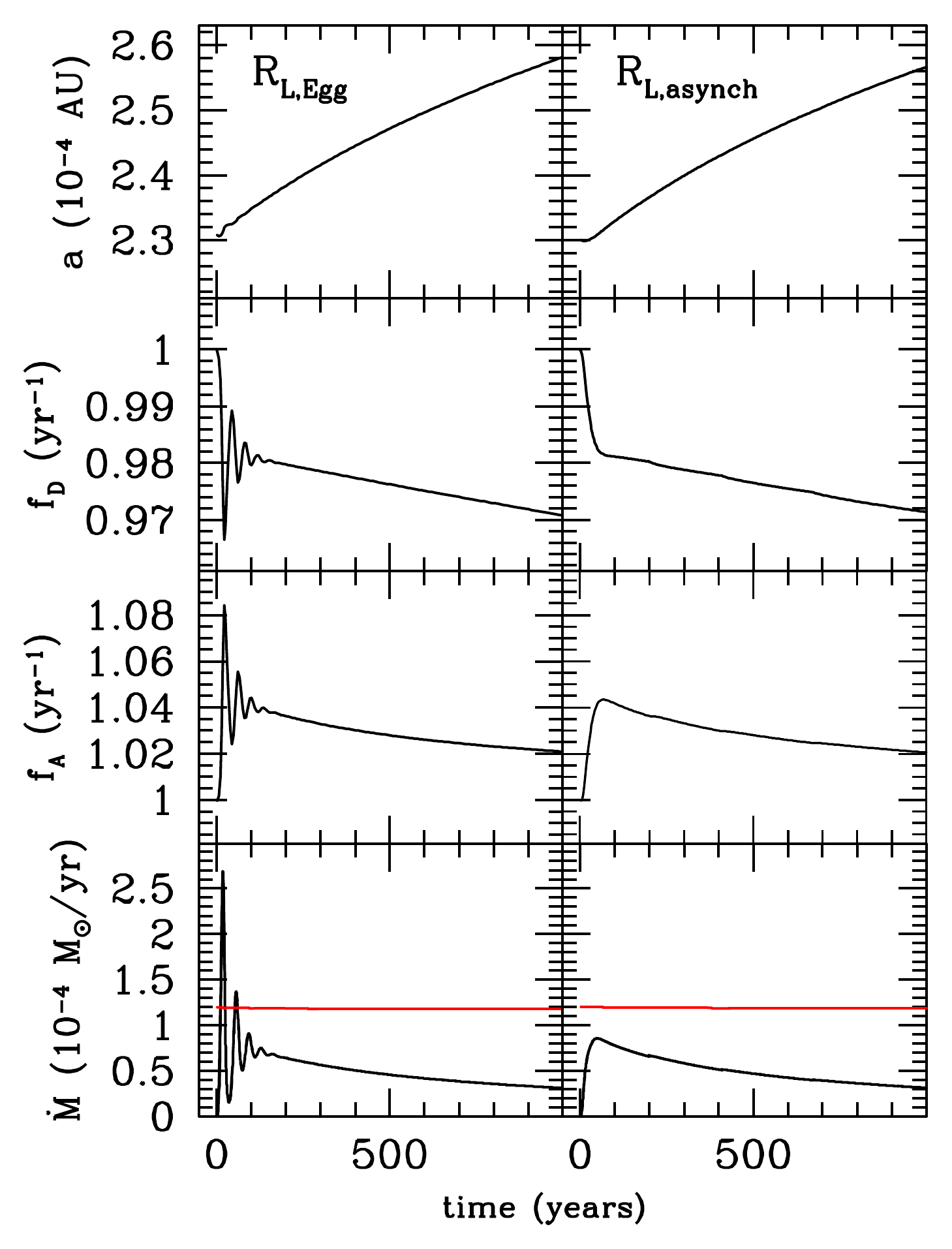}
\caption{ \footnotesize Evolution of orbital parameters for $\tau$ = 10 years for initial masses, $M_D = 0.38~M_{\odot}$ and $M_A = 0.8~M_{\odot}$. First 1000 years of evolution with solutions obtained using $R_{L,{\rm Egg}}$ for the calculation of the Roche lobe on left and using $R_{L,asynch}$ on right. The red lines show the super-Eddington accretion rate.}
\end{figure}

\begin{figure} [t!]
\plotone{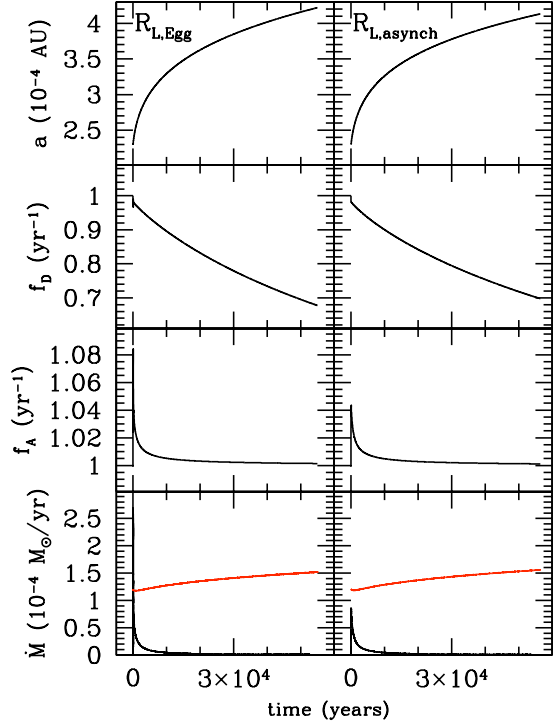}
\caption{ \footnotesize Full evolution in time of orbital parameter for $\tau$ = 10 years with $M_D = 0.38~M_{\odot}$and $M_A = 0.8~M_{\odot}$. Compare to Figure 8 which shows only the first 1000 years of evolution for this particular system.}
\end{figure}

Figure 8 compares the evolution of a system with an initial donor mass 
of $0.38\,M_{\odot}$ and an initial accretor mass of $0.8\,M_{\odot}$ 
with $\tau = 10\,$years for the 
first 1000 years of evolution. Figure 9 shows the full evolution of this 
system, which evolved out to approximately $6 \times 10^4$ years before reaching a phase of disc accretion. As in Figure 7, the 
left panel shows the evolution using $R_{L,{\rm Egg}}$ and the right panel shows the evolution using $R_{L,asynch}$. As can be seen in Figures 3 
and 5, this system is stable under the $R_{L,asynch}$ calculation and 
super-Eddington under the $R_{L,{\rm Egg}}$ calculation. Unlike the 
$10^{15}$ year time-scale, for the 10 year time-scale, the orbital 
parameters follow approximately the same path in both the $R_{L,{\rm 
Egg}}$ and $R_{L,asynch}$ cases, with the exception being the 
oscillations seen when using $R_{L,Egg}$.  We discuss these oscillations in 
more detail below. 

\subsection{Oscillations of orbital parameters}

As seen in Figure 8, using $R_{L,asynch}$ to 
calculate the Roche lobe damps the oscillations 
observed when using the $R_{L,{\rm Egg}}$. In both cases, the mass 
transfer rate increases initially, which causes the accretor to spin up 
and the donor to spin down. In the $R_{L,asynch}$ case, this causes the 
mass transfer rate to increase at a slower rate relative to the 
$R_{L,{\rm Egg}}$ case. \footnote{As described previously, as 
$f_D$ decreases, $\mathcal{A}$ decreases, which causes the Roche lobe to 
increase and reduces the mass transfer rate.} As discussed above, when 
using $R_{L,{\rm Egg}}$, the mass transfer rate will continue to increase 
until tides have transferred sufficient angular momentum from the spin 
of the components to the orbit to widen the orbit and reduce the mass 
transfer rate. Once this occurs, the mass transfer rate is reduced and tides continue to redistribute 
angular momentum between the component spins 
and the orbit (in this case the donor is spun up 
while the accretor is spun down). But again, because changes in the 
donor spin are not accounted for when using $R_{L,Egg}$, the mass transfer rate 
will again ``over shoot" and the oscillations continue. The 
dependence of the Roche lobe radius upon the donor spin when using $R_{L,asynch}$ slows the changes in the mass transfer rate and damps the 
oscillations.

\begin{figure} [t!]
\plotone{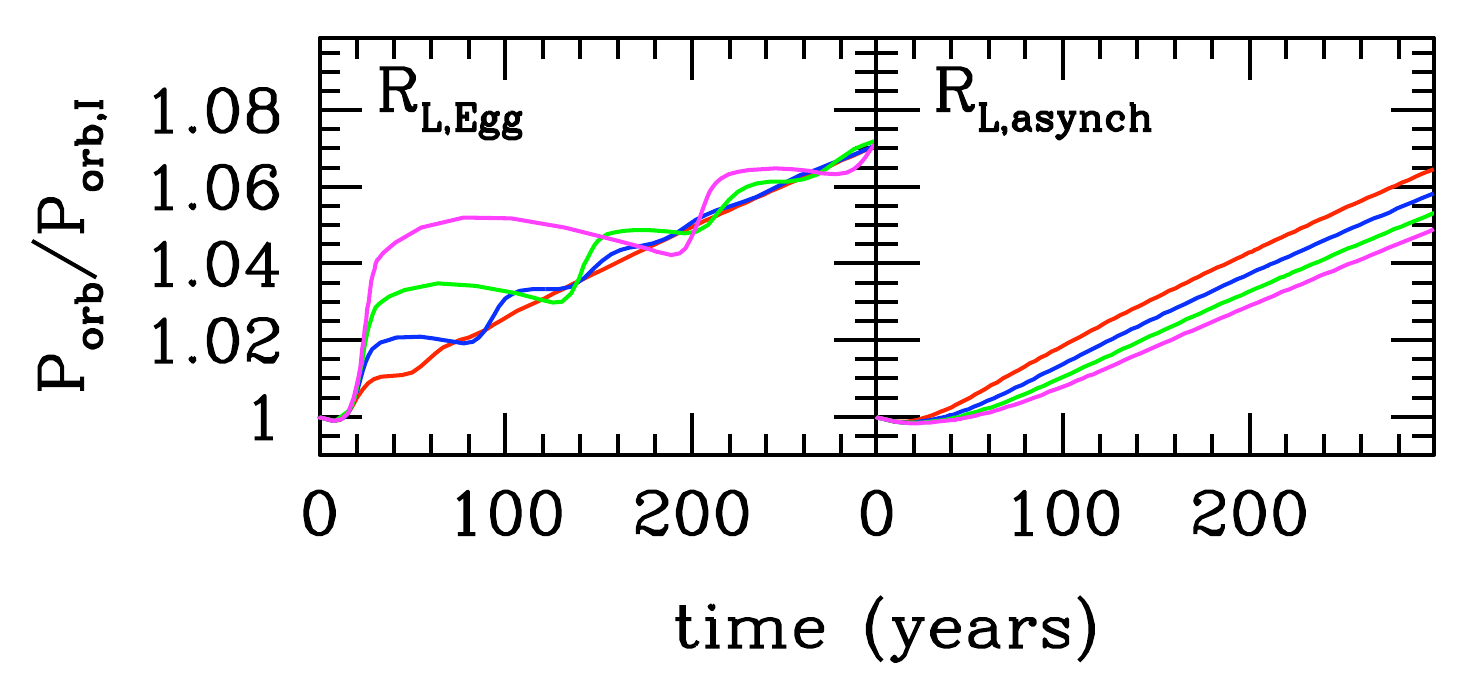}
\caption{\label{fig:Figure5} \footnotesize Evolution of the orbital period, $P_{orb}$, compared to its initial orbital period, $P_{orb,I}$, for a 
system with $M_{D} = 0.38\,M_{\odot}$ and $M_{A} = 0.8\,M_{\odot}$ for 
various synchronization time-scales at contact. Red, blue, and magenta lines represent the orbital periods for 
synchronization timescales of 20, 30, and 40 years, respectively. 
The left-hand panel shows the evolution using $R_{L,{\rm Egg}}$ and the 
right-hand panel shows the evolution using $R_{L,asynch}$.}
\end{figure} 

Although Figure 8 shows that the short-term evolution of this particular 
system differs substantially between the $R_{L,{\rm Egg}}$ and 
$R_{L,asynch}$ Roche lobe calculations, Figure 9 shows that the 
long-term evolution of this system is very similar. Although the oscillations do not 
impact the long-term evolution of many of the systems, the oscillations 
do impact whether a system is categorized as sub-Eddington or 
super-Eddington or whether it reaches unstable levels of mass 
transfer.  In the case of the system illustrated in Figures 8 and 9, 
the oscillations cause the system briefly to become 
super-Eddington, while the system would remain 
sub-Eddington in the absence of oscillations.

In general, as the mass transfer rates spike initially under 
the $R_{L,{\rm Egg}}$ calculation before ultimately settling to the 
$R_{L,asynch}$ value, it may reach super-Eddington or unstable levels, while, in the $R_{L,asynch}$ case, the mass transfer 
rate may remain stable instead of super-Eddington, or 
super-Eddington instead of unstable. This explains the lack of unstable 
black systems in Figure 5 compared to Figure 3. For
the systems shown in Figure 3, the oscillations 
in the orbital parameters arising from the use of $R_{L,Egg}$ may cause the mass transfer 
rate to reach artificially high values before the oscillations relax, 
leading to the large population of black systems at the top of the 
figure. Meanwhile, since using $R_{L,asynch}$ damps the oscillations, 
the mass transfer rate is able to stay below the limit for stability of 
0.01 $M_{\odot}/$yr.

\begin{figure} [t!] 
\plotone{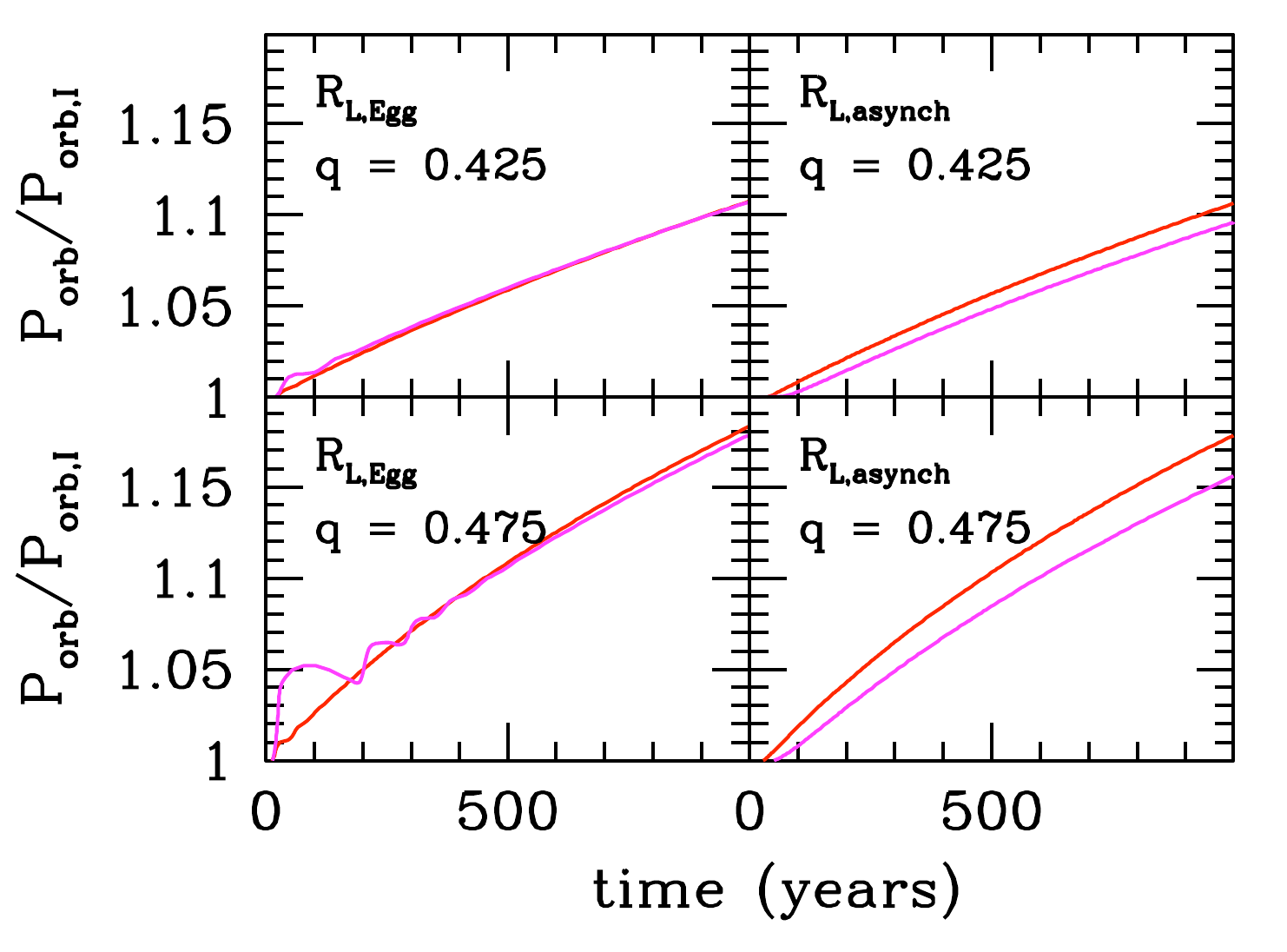}
\caption{\label{fig:Figure6} \footnotesize Evolution of the orbital period for two different mass ratios. The upper graphs show the evolution for an initial mass ratio of q = 0.425 ($M_{D} = 0.34~M_{\odot}$ and $M_{A} = 0.8~M_{\odot}$) and the lower graphs show an initial mass ratio of q=0.475 ($M_{D} = 0.38~M_{\odot}$ and $M_{A} = 0.8~M_{\odot}$). The left-hand panel shows evolution using $R_{L,{\rm Egg}}$ and the righ-hand panel shows evolution using $R_{L,asynch}$. As in Figure 10, the red plots show evolution for an initial synchronization time-scale of $\tau = 10$ yrs and the magenta plots show the evolution for $\tau = 40$ yrs.}
\end{figure}

Both Marsh2004 and Gokhale2007 noted the presence of oscillations in the orbital parameters 
near the boundary between stable and unstable systems. These analyses observed that, for such systems, small 
changes in the synchronization time-scales and in the mass ratio will 
cause changes in the frequency and amplitude of the oscillations. 
Figures 10 and 11 show the evolution of the orbital period for various 
synchronization time-scales and mass ratios, respectively, using both $R_{L,{\rm Egg}}$ and $R_{L,asynch}$ to calculate the size of the Roche lobe. 
These figures confirm the observations of Marsh2004 and Gokhale2007 that the amplitude and 
frequency of the oscillations is dependent upon both the mass ratio and 
the synchronization time-scale.

The left-hand panel of Figure 10 illustrates that the amplitude of the 
oscillations increases with increasing initial 
synchronization time-scale, while the frequency of the oscillations 
decreases.  These 
oscillations can be explained as follows: Once mass transfer begins, if 
the system is unstable or nearly unstable, the time-scale for changes in the mass transfer rate becomes greater than 
the time-scale at which 
the semi-major axis increases due to tides. As a result, $\dot M_D$ 
increases rapidly and the accretor spin 
increases significantly. The result is a large asynchronicity between the accretor, the orbit, and the donor. 
As the orbital separation increases due to the effect of tides, 
$\dot M_D$ decreases, until the timescale for changes in the 
mass transfer rate becomes less than the time-scale at which the semi-major axis increases due to 
tides. At this point, tides are able to redistribute angular momentum 
from the spin of the accretor back into the orbit. If enough angular 
momentum is transferred back into the orbit, the donor becomes detached. Once tides 
effectively re-synchronize the spins of the stars with the orbit, 
gravitational wave losses act to bring the binary back into contact, and 
the entire cycle repeats.
 
The right-hand panel of Figure 10 demonstrates that, for this particular 
mass ratio, the oscillations are not 
present when using $R_{L,asynch}$ to calculate the size of the Roche 
lobe. Furthermore, the right-hand panel 
illustrates that, in the absence of oscillations, the orbital period, and therefore, the semi-major axis increases more rapidly for lower $\tau$ values of the synchronization timescale. 
For lower $\tau$ values, the tidal coupling is stronger, which means spin angular 
momentum is able to be re-distributed more efficiently, which allows the semi-major axis to remain larger compared to the semi-major axis 
of systems evolved using higher values of $\tau$.

Figure 11 demonstrates the effect of mass ratio upon the amplitude of 
the oscillations. As Gokhale2007 observes, we see that for a fixed 
$\tau$ value, a lower mass ratio (top two graphs of 
Figure 11) 
yields oscillations of lower amplitude compared to a higher mass ratio 
(lower two graphs of Figure 11). In Figure 3, we see that switching from a mass ratio of 0.475 to 0.425 moves us away from the 
boundary which separates super-Eddington systems from stable systems. 
This demonstrates that, in general, as we move away from the stability 
boundary, the oscillations are reduced.

The two right-hand panels of Figure 11 show the evolution using 
$R_{L,asynch}$ for the roche lobe. This again demonstrates that 
oscillations are not present when the Roche lobe depends on the asynchronicity 
between the donor and orbit.

\section{Conclusions}

We have studied the long-term evolution of DWD binary systems undergoing direct-impact mass transfer, including the effects due to mass transfer, gravitational radiation, and tidal forces arising from asynchronicity between the donor, accretor, and the orbit. We implemented the ballistic mass-transfer treatment developed in \citep{Sepinsky2010} to calculate the changes to orbital parameters during direct-impact mass transfer. By implementing this method, we found that the number of stable DWD systems increased for both weak and strong tidal coupling compared with the results of Marsh2004 and Gokhale2007, as shown in Figures 2 and 3.

For the first time, we also account for the modification of the Roche-lobe size due to the asynchronicity of the donor. As a result, we find that the number of stable systems increases, particularly for the case of strong tidal coupling, as shown in Figures 5 and 6. When not accounting for the asynchronicity effects on the Roche-lobe size, we reproduce the oscillations in the orbital parameters first noted by Gokhale2007. We found that, when the size of the Roche lobe was permitted to vary with donor asynchronicity, these oscillations were dampened, as shown in Figure 8. We conclude that the oscillations created when using the Eggleton approximation for the Roche lobe calculation create artificially high mass transfer rates which leads to an artificially high number of super-Eddington and unstable systems. By eliminating the oscillations, our treatment yields a higher number of sub-Eddington and stable systems.

We expect systems which are stable throughout our calculations (red systems in Figures 2, 3, 5, and 6) to be AM CVn progenitors. As a result of the increase in stable systems shown here compared to previous analyses, we conclude that DWD evolution may be a more likely avenue for the creation of AM CVn than previously expected.

In future analyses, we intend to investigate the case of initially eccentric binaries with asynchronous component stars, as we expect there may be a significant population of 
DWDs that have not had time to circularize and synchronize by the time 
mass transfer occurs \citep{Willems2007}. Additionally, we intend to implement a treatment of disc accretion so that we can track the evolution of the DWD systems through all types of accretion processes.

\section{Acknowledgements}

KK and VK acknowledge support from Northwestern University that made this project possible. VK is also grateful for the hospitality of the Aspen Center for Physics. This work used computing resources at CIERA funded by NSF PHY-1126812.

\end{document}